\documentclass{dcds-b}
\usepackage{amsmath}
  \usepackage{graphics} 
  \usepackage{epsfig} 
 \usepackage[colorlinks=true]{hyperref}
\hypersetup{urlcolor=blue, citecolor=red}

\def\currentvolume{12}
 \def\currentissue{3}
  \def\currentyear{2009}
   \def\currentmonth{October}
    \def\ppages{1--XX}
     \def\DOI{10.3934/dcdsb.2009.12.xx}

\newtheorem{theorem}{Theorem}[section]

\theoremstyle{definition}
\newtheorem{definition}[theorem]{Definition}
\newtheorem{remark}{Remark}

\def\p{\partial}

\def\a{\alpha}
\def\b{\beta}
\def\g{\gamma}

\def\o{\omega}
\def\e{\varepsilon}
\def\wt{\widetilde}

\def\be{\begin{equation}}       \def\ba{\begin{array}}

\def\ee{\end{equation}}         \def\ea{\end{array}}

\def\bea {\begin{eqnarray}}      \def\eea {\end{eqnarray}}

\def\bean{\begin{eqnarray*}}    \def\eean{\end{eqnarray*}}

           \def\ph{\varphi}

\def\RA {\ \Rightarrow\ }         

\def\qed   {\vrule height0.6em width0.3em depth0pt}

\def\<{\langle} \def\({\left(}  \def\>{\rangle} \def\){\right)}

\newtheorem{exi}{Example}

\def\a{\alpha}
\def\e{\varepsilon}
\def\wt{\widetilde}

\title[Nonlinear resonances of Water Waves]
      {Nonlinear resonances of Water Waves}

\author[Elena Kartashova]{}


 \email{lena@risc.uni-linz.ac.at}

\thanks{The author is supported by FWF grant P20164-N18}

\begin{document}
\maketitle

\centerline{\scshape Elena Kartashova }
\medskip
{\footnotesize
 \centerline{RISC, J. Kepler University}
   \centerline{Altenbergerstr. 69}
   \centerline{ Linz, A-4040, Austria}
} 

\bigskip

 \centerline{(Communicated by the associate editor name)}

\begin{abstract}
In the last fifteen years great progress has been made in the
understanding of nonlinear resonance dynamics of water waves.
Notions of scale- and angle-resonances have been introduced, new
type of energy cascade due to nonlinear resonances in the gravity water
waves has been discovered, conception of a resonance cluster has
been much and successfully employed, a novel model of laminated wave
turbulence has been developed, etc. etc. Two milestones in this area
of research have to be mentioned: a) development of the $q$-class
method which is effective for computing integer points on
resonance manifolds, and b) construction of marked planar graphs, instead of classical resonance curves, representing simultaneously all resonance clusters in a finite spectral domain, together with their dynamical systems. Among them, new integrable dynamical systems have
been found that can be used for explaining  numerical and
laboratory results. The aim of this paper is to give a brief
overview of our current knowledge about nonlinear resonances among
water waves, and finally to formulate the three most important open problems.
\end{abstract}

\section{Exposition}

In this paper we will try to present a major part of known analytical,
numerical and laboratory results on nonlinear resonances among water
waves, in as strict mathematical language as possible. This is not a
simple task due to the three-fold problem: 1) there is no strict
definition of a wave; 2) there is no general agreement about the
types of waves which should be called water waves; 3) the notions of
resonance in physics and mathematics are different. Let us go through
all these points one by one, regarding for concreteness
2\textbf{D}-wavevectors.

\textbf{First},
the simplest possible understanding of a (propagating) wave as a Fourier harmonics%
\be \label{Fourier harmonics} A_\mathbf{k} \exp {i(
{\mathbf{k}}\cdot{ \mathbf{x}} -
\o\, t)}\ee %
 is obviously too simplified and does not
include normal modes which are due to boundary conditions. Here
$\mathbf{x}=(x_1,\, x_2\,)$ and time $t$ are space and time
variables correspondingly, $\o=\o(\mathbf{k})$ is dispersion
function and $\mathbf{k}$ is wavevector. For instance, the normal mode
of oceanic planetary
waves (that are due to the Earth rotation) with zero boundary conditions in a
finite box $[0,L_{x_1}]\times[0,L_{x_2}]$ reads \cite{Ped87} %
\be \label{rectangular_mode}|A_\mathbf{k}|
\sin\!\Big(\pi\frac{mx_1}{L_{x_1}}\Big)
\sin\!\Big(\pi\frac{nx_2}{L_{x_2}}\Big)\, \sin\!
\Big(\frac{\b}{2\o}x_1+\o t\Big ). \ee %
Boundary conditions on a sphere or a circle lead to even more
complicated forms of the normal mode including special functions
\cite{Ped87}. "We seem to be left at present with the looser idea
that whenever oscillations in space are coupled with oscillation in
time through a dispersion relation, we expect the typical effects of
dispersive waves" \cite{Whi99}. So, from now on we assume that
dispersion function $\o(\mathbf{k})$ defines the type of a wave, for
instance, $\o \sim |\mathbf{k}|^{3/2}$ corresponds to capillary
waves in a rectangular box with periodic boundary conditions, $\o
\sim |\mathbf{k}|^{-1}$ corresponds to oceanic planetary waves in a
rectangular box with zero boundary conditions, etc. Sign "$\sim$"
instead of "$=$" in definitions of dispersion function means that
some constant is omitted.

\textbf{Second}, capillary, gravity-capillary, gravity surface,
oceanic planetary and freak waves, tsunami, internal waves in
rotating fluid, etc., etc. can be observed in water, and they all
can be -- and sometimes are  -- called water waves. This is, of
course, too extensive a definition that does not allow any
reasonable overview, neither in the form of a paper nor of a book.
An example of another extreme can be given by \cite{KMV02} where
only waves on the surface of deep water are called "water waves". In
this paper we choose to follow the classical conception of
 water waves \cite{Whi99}, as defined by
dispersion function%
\be \label{disp_general} \o^2=(g \, k \tanh k \,
d)(1+\frac{\sigma}{\rho g} k^2), \quad \mbox{with} \quad
k=|\mathbf{k}|\ee%
and $ g, \, d, \, \sigma, \, \rho $ being  the gravitational
constant of acceleration, the average water depth, the coefficient
of surface tension and density correspondingly. In this paper we
will concentrate on, though not restrict ourselves completely by, the
three limiting cases of (\ref{disp_general}) for water waves, with
$\rho=1$ and $k \, d \rightarrow \infty $, which are important
for numerous applications:%
\bea %
&1.&  \quad \mbox{capillary waves,} \quad \lambda_{ch} \ll 2 \, \,
\mbox{cm}: \quad
 \o^2 =\sigma k^{3},  \label{CapDeepWat} \\
&2.& \quad \mbox{gravity waves,} \quad \lambda_{ch} \gg 2 \, \,
\mbox{cm}: \quad
 \o^2 =g k, \label{GrDeepWat}  \\
&3.& \quad \mbox{gravity-capillary waves,} \quad \lambda_{ch} \sim 2
\, \, \mbox{cm}: \quad
\o^2 = g \, k + \sigma k^3,\label{GrCap}%
\eea%
where $\lambda_{ch}$ denotes characteristic wave length for each
approximation.

\textbf{Third}, the notion of resonance appears in mathematics in
the theory of Poincar\'e normal forms, as a condition of
linearization of a nonlinear ODE or a system of nonlinear ODEs
\cite{Mu03}, in the form%
 \be \label{ResMath} p_1\o_1+ \cdots
+p_s\o_s+p_{s+1}\o_{s+1}=0 \nonumber \ee%
 where $\o_j$ is notation for $\o(\mathbf{k}_j)$ and $p_1, \, p_2, \, \dots, \,
 p_{s+1} \in \mathbb{Z} $ are integers. The physical notion of resonance
 has been first introduced by Galileo Galilei \cite{Galileo} who
 studied oscillations of a pendulum under the action of a small external
 force. In modern language, the corresponding equation reads
$x_{tt} + r^2 x = \e \exp{(i f t)}$ with $ 0< \e \ll 1,$ and the
oscillation amplitude grows linearly with time if eigenfrequency of
the system $\o$ coincides with the frequency of the driving force $f.$
 The analogy with the three-wave resonance is now obvious:
 two Fourier harmonics of the form (\ref{Fourier harmonics}) with
 wavevectors $\mathbf{k}_1, \, \mathbf{k}_2,$
frequencies $\o_1, \, \o_2$ and amplitudes $A_1, \, A_2$ will give
resonance with harmonic %
\be \label{Fourier harmonics_res} A_{3} \exp {[i( {\mathbf{k}_1 \pm
\mathbf{k}_2})\cdot{ \mathbf{x}} -
(\o_1 \pm \o_2)\, t]}, \nonumber \ee %
which yields 3-wave resonance conditions in the form%
 \be \label{3res}%
\o_1+\o_2=\o_3, \quad \mathbf{k}_1+\mathbf{k}_2=\mathbf{k}_3.
\ee%
Similarly, 4-wave resonance conditions read%
 \be \label{4res}%
\o_1+\o_2=\o_3+\o_4,\quad
\mathbf{k}_1+\mathbf{k}_2=\mathbf{k}_3+\mathbf{k}_4.
\ee%
In the case of periodic or zero boundary conditions, solutions of
(\ref{3res}),(\ref{4res}) have to be found in integers, i.e. for
$\mathbf{k}_j=(m_j, n_j)$ with $m_j, \, n_j \in \mathbb{Z}.$
Solutions of (\ref{3res}) and (\ref{4res}) are called triads and
quartets correspondingly, they are defined by the geometry  of the
wave system and do not depend on time, that is, they
 describe \textbf{kinematics} of a wave system.
\textbf{Dynamics} of a resonant triad is covered by%

\be \label{3prim}  i\, \dot{A}_1 = {V^3_{12}}^{\ast}
A_2^{\ast}A_3,\quad i\, \dot{A}_2 = {V^3_{12}}^{\ast} A_1^{\ast}
A_3, \quad
i\, \dot{A}_3 = V^3_{12} A_1 A_2,%
 \ee 
 where  $A_j$ are amplitudes of three waves satisfying
 (\ref{3res}) and $\dot{A}=\frac{d}{d T}{A}$. Dynamical system for a resonant quartet reads
  \bea \label{4prim}
  i\, \dot{A}_1 =   T^{12}_{34} A_2^{\ast}A_3A_4 + (\Omega_1 - \o_1)A_1\,,\quad
 i\, \dot{A}_2 =   T^{12}_{34} A_1^{\ast}A_3A_4+ (\Omega_2 - \o_2)A_2 \,, \nonumber \\
 i\, \dot{A}_3 =   {T^{12}_{34}}^{\ast} A_4^{\ast}A_1A_2+ (\Omega_3 - \o_3)A_3\,,
 \quad
 i\, \dot{A}_4 =  {T^{12}_{34}}^{\ast} A_3^{\ast}A_1A_2 + (\Omega_4 - \o_4)A_4\ .%
  \eea 
where $\Omega_j$ are so-called "Stocks-corrected" frequencies \cite{SS05}.  They are often omitted in the literature while suitable renormalization of $\o$-s yields more usual form of (\ref{4prim}), without terms $(\Omega_j - \o_j)A_j$ included (e.g. \cite{LKC09}, for one-dimensional quartets).  The general form of (\ref{3prim}), (\ref{4prim}) can be deduced in
the frame of the Hamiltonian formalism, \cite{ZLF92}, and
\textbf{does not depend} on the type of waves. The only difference
between various wave types is hidden in the form of coupling
coefficients $V^3_{12}$ and $T^{12}_{34}$; their expressions for
water waves are given in the Table 1. Resonant triads and quartets may form \emph{resonance clusters},  for instance when one  resonant mode $(m,n)$ is
part of a few solutions of (\ref{3res}) or (\ref{4res}). Obviously, triads and quartets are the minimal resonance clusters in three- and four-wave systems correspondingly; they are called \emph{primary clusters}.

\begin{remark}
To use the Hamiltonian formalism for describing these wave systems,
some small parameter $0 < \e \ll 1$ has to be introduced, which allows the
application of variational or multi-scale methods. These yield, in the
the simplest case, two time-scales - \emph{fast time}, $t$, for
the linear Fourier harmonic (\ref{Fourier harmonics}), and \emph{slow
time}, $T=\e \, t $, for the amplitudes of the nonlinear resonant mode
(\ref{3prim}). In water wave systems, wave steepness is usually taken as
a small parameter, $\e \sim 0.1$.
\end{remark}

\section{Kinematics}

\subsection{Exact resonances}

Computing integer solutions of resonance conditions (\ref{3res}),
(\ref{4res}) for dispersion functions
(\ref{CapDeepWat})-(\ref{GrDeepWat}) is equivalent to finding
integer points on a resonant manifold of a high degree in many
variables. Loosely speaking, the problem is equivalent to
Hilbert's Tenth Problem \cite{Hi02}, which is proven to be
unsolvable in the general case \cite{Mat70}, and even for the case of two
variables only partial results are known. We are interested in
solutions of equations (\ref{3res}), (\ref{4res}) with 6 and 8
variables correspondingly. One could hope that numerical search for
solutions might be a reasonable possibility but it is not. Full
search for multivariate problems in integers consumes exponentially
more time with the size of the domain to be explored and for each additional variable.
 For (\ref{4res})  in the domain $ |m|, \, |n|
\sim 1000$ this would imply some $10^{24}$ tries. Still
worse, the equations include radicals and straightforward
transformation to a purely integer form would lead to operations
with huge integer numbers - for the said domain, of the order of
$10^{120}$. All these reasons make a quest for effective algorithms
unavoidable.

A special method for solving systems of the form (\ref{3res}),
(\ref{4res}) has been presented in \cite{K06-3} for gravity waves
with dispersion function (\ref{GrDeepWat}). Details of its
implementation and numerical results for various irrational
dispersion functions can be found in \cite{KK06-1,KK06-2}. The
method is called $q$-class method or $q$-class decomposition. Its
idea comes from the linear independence of a finite number of some
algebraic numbers. Below we give the general definition of the
$q$-class and show how to use it for finding solutions of
(\ref{3res}), (\ref{4res}).
\bigskip
\begin{definition} For a given $c \in \mathbb{N}, c \neq 0, 1, -1$ consider the set of
algebraic numbers $R_c= {±k^{1/c}, k \in \mathbb{N}}$. Any such
number $k_c$ has a unique representation
$$
    k_c = \g q^{1/c} , \g \in \mathbb{Z} \quad \mbox{with} \quad q=p_1^{e_1} p_2^{e_2} ... p_n^{e_n},
$$
and $p_1, ... p_n$ being all different primes and the powers $e_1,
...e_n \in \mathbb{N} $ are all smaller than $c$. The set of numbers
from $R_c$ having the same $q$ is called $q$-class $Cl_q$. The
number $q$ is called class index.
\end{definition}
\bigskip
Now we can decompose the computational domain into $q$-classes and
search for solutions in each class separately.
 The computational advantage of using $q$-class
decomposition is huge. For instance, for dispersion function
(\ref{GrDeepWat}) and resonance conditions (\ref{4res})
straightforward computation takes 3 days with Pentium-4 in the
spectral domain $|m|, |n|\le 128$; the $q$-class decomposition
produces all solutions for $|m|, |n|\le 1000$ in about 15 minutes
with Pentium-3.

\subsubsection{Capillary waves, $\o^2 = \sigma k^3,$ three-wave resonances.}%

In this case, we take $c=2$ in the definition above and represent the
norm $k_j$ as%
 \be \label{Fermat1}k_j=\g_j \sqrt{q_j}, \quad \g_j, \,q_j \in \mathbb{N}, \quad q_j \quad \mbox{is square-free} \ee%
The presentation (\ref{Fermat1}) is unique for each $k_j$ and it
follows that%
\be \label{Fermat2} \o_1+\o_2=\o_3 \quad \RA \quad \g_1^3
q_1\sqrt{q_1}+\g_2^3 q_2\sqrt{q_2}=\g_3^3
q_3\sqrt{q_3}.\ee%
The necessary condition of the existence of integer solutions of
(\ref{Fermat2}) is $ q_1 = q_2 = q_3,$ say $q_j=q, \, \forall
j=1,2,3,$ (see \cite{AMS98}),  and this allows to reduce
(\ref{Fermat2}) to a particular case of Fermat's Last Theorem,
$\g_1^3 + \g_2^2 = \g_3^3, $ i.e. there are no resonances among
capillary waves with dispersion function (\ref{CapDeepWat}).

The case of capillary waves is unique. Usually, a three-wave system
possesses a  number of resonances divided into small resonance clusters
\cite{K94}. Mostly, resonance clusters are isolated triads or groups
of two connected triads ($70-100 \%\%$ of all clusters, depending on
the size of the spectral domain). For instance, the same $q$-class
construction used for ocean planetary waves with dispersion
function $\o \sim k^{-1/2}$ and resonance conditions (\ref{3res}),
turns the first equation of (\ref{3res}) into $\g_2 \, \g_3 + \g_1 \,
\g_3 = \g_1 \, \g_2,$ in which case solutions do exist
\cite{KK06-1}.

At the end of this section we would like to mention quite recent
progress on the resonances of capillary waves reported in
\cite{CK09}. Dispersion function has
been taken in the form%
 \be \label{DispCap} \omega = \frac{\Omega}{2}
+ \frac{1}{2} \sqrt{\Omega^2 + 4 k^3\sigma }, \ee %
where $\Omega $ is non-zero constant vorticity. The main results
are as follows: 1) wave trains in flows with constant
non-zero vorticity are possible only for two-dimensional flows; 2)
only positive constant vorticities can trigger the appearance of
three-wave resonances; 3) the number of positive constant
vorticities which do trigger a resonance is countable; 4) the
magnitude of positive constant vorticity triggering a resonance
can not be too small. Of course, these results are only valid for waves with small amplitude \cite{CK09,CSS,CS,W3} (cf. Remark 1 above).

\subsubsection{Gravity-capillary waves,
$\o^2 = g \, k + \sigma k^3,$ three-wave resonances.}%

Construction of $q$-classes is this case differs from
(\ref{Fermat1}) and is more intricate \cite{KK09}, due to the fact
that dispersion function contains two irrationalities and is not homogenous on $k,$ i.e. coefficients $g$ and $\sigma$ do not disappear.
For numerical simulations, we used
 $g=981$ and two values of $\sigma$: $\sigma=74$ and $\sigma=75$, for water with temperature $18^{\circ}$C and $8^{\circ}$C correspondingly.
Numerical simulations have been performed, seeking solutions $\o_1 + \o_2 = \o_3(1+\e)$, for various $\e,$ $0 \le |\e| \le 10^{-4}$. Spectral domain $-100 \le m, \, n \le 100$ has been investigated. For both cases, $\sigma=74$ and $\sigma=75$, only isolated triads have been found (for $\e=0$): 24 and 16 triads correspondingly; no triad appears simultaneously in both lists. The resonance structure is much richer for $\e>0,$ resonance clusters of two to six triads appearing already for $\e= 10^{-7}$, while for $\e= 10^{-4}$ the overall number of resonances is about $10^6,$ and more than $90\%$ of all resonant triads coincide for
$\sigma=74$ and $\sigma=75$.

\subsubsection{Gravity waves, $\o^2 =g k,$ four-wave resonances.}

In this case, $c=4$ in the definition of the $q$-class, i.e.
$k=(m^2+n^2)^{1/4}=\g q^{1/4}$, where $q$ does not contain fourth
degrees in its representation as a product of different primes in corresponding
powers \cite{K06-3}. It can be proven that only two cases are
possible: 1) all 4 wavevectors, forming a solution of (\ref{4res}),
belong to one class $Cl_q$, and 2) two wavevectors belong to
$Cl_{q_1}$ and two others belong to $Cl_{q_2}$ with $q_1 \neq q_2.$
Again, this is only a necessary, not a sufficient condition for the
existence of a solution.

From physical point of view, the main difference between nonlinear
resonances in three- and four-wave systems can be formulated as
follows. Any three-wave resonance generates a \emph{new frequency}
and, correspondingly, new wave length, i.e. any three-wave resonance
contributes to the energy transfer over the scales. This is not true
in four-wave systems. Indeed, let us regard one specific solution of
(\ref{4res}): $ \mathbf{k}_1 = ( -64, -16),\, \mathbf{k}_2 = (4,16),
\,  \mathbf{k}_3 = (-64,16), \,  \mathbf{k}_4 = (4 ,-16).$ This
solution is not trivial - all wavevectors are different, but $k_1 =
k_3$ and $k_2 = k_4,$ i.e. no new wave length appears due to this
kind of resonances. Being regarded in $\mathbf{k}$-space, these
resonances perform the energy transfer not over the scales but over
the angle, forming two circles. Correspondingly, two types of
resonances are singled out, with different dynamics - \emph{angle-}
and \emph{scale-resonance} \cite{K07}.

The fact of utmost importance is that these two types of
resonances should not be studied separately, because \emph{mixed
cascades} also exist. For instance, the wavevector (119,120) takes
part in one scale-resonance and 12 angle-resonance \cite{K07}.
Complete study of all the resonances in the spectral domain
$|m_i|,|n_i| \le 1000$ can be found in \cite{KNR08}. In this domain,
there are more than 600.000.000 exact resonances, among them only
230.464 angle-resonances, among which 213.760 resonances are
formed by four collinear wavevectors. As it is shown in
\cite{DLZ95}, the coupling coefficient (see Table 1) in this case is
equal to zero, i.e. these resonances do not affect dynamics. Some parametric series of solutions for the resonance conditions
(\ref{4res}) are known, for instance, for any given wavevector
$(m,n),$ a 3-parametric series of angle-resonances is known:
 \be \label{Angle1}
 {\bf k}_1=(m,n), \ {\bf k}_2=(t,-n), \ {\bf k}_3=(m,-n), \ {\bf
k}_4=(t,n), \quad t=0,\pm 1, \pm2, \ldots .
 \ee
Series of scale-resonances called \emph{tridents} has been first
found in \cite{LNP06} for quartets of the form:%
 \be \label{trident}
{\bf k}_1=(a,0),\  {\bf k}_2=(-b,0),\  {\bf k}_3=(c,d),\ {\bf
k}_4=(c,-d)\,, \ee
where $ a=(s^2+t^2+st)^2, \ b=(s^2+t^2-st)^2, \ c=2st(s^2+t^2), \
d=s^4-t^4\,, s,t \in \mathbb{N}. $ This series is of special interest for
studying large-scale gravity waves: all scale-resonances in the
spectral domain, say, $k \le 100$ are of this form \cite{KNR08} and
therefore can be found analytically. Analytical series are very
helpful not only for computing resonance quartets and cluster
structure but also for investigating the asymptotic behavior of
coupling coefficients. More details can be found in \cite{KNR08}.

\begin{remark}
The results reported in this section were possible to obtain because all three dispersion functions (\ref{CapDeepWat})-(\ref{GrDeepWat}) are irrational and we could make use of the $q$-class method. Interestingly enough, similar idea  can also be worked out into a computational method  for a transcendental dispersion function, say, $\o^2 =g \, k \tanh k\, d.$ Indeed, using standard presentation
$\tanh x= [\exp{x}-\exp{(-x)}]/[\exp{x}+\exp{(-x)}],$ one can rewrite (\ref{4res}) as a combination of different exponents with polynomial coefficients depending on $k$ and $d$ (see \cite{K92}, Discussion, p.52). Afterwards appropriate theorems about linear independence of exponents over algebraic numbers can be used \cite{Ba75}. This procedure can be used for any transcendental dispersion function which has a representation as a rational function of different exponents.
\end{remark}

\subsection{Topological structure of the complete resonance set}

In 1950-1960th the usual way to represent
nonlinear resonances was  to construct \emph{a resonance curve} that
is the locus of pairs of wavevectors interacting resonantly with a
given wavevector. A locus might be an ellipse, be shaped like an
hour-glass, can degenerate into a pair of lines, etc. Eventually,
the characteristic form of the locus is useful to know before
planning some laboratory experiments: in \cite{Ph67} a resonance
curve has been constructed for a special type of resonant quartets
of gravity waves, in which two wavevectors coincide. A typical
resonance curve for gravity-capillary waves can be found in
\cite{Ph74}. The main drawback of the resonance curve presentation
is that it allows to visualize -- at most -- only \textbf{a part of
the resonance set}. And of course, without the $q$-class method, there
has been no constructive way to establish whether any
integer points exist on these curves.

A novel graphical presentation of \textbf{2D}-wave resonances as a
planar graph, \cite{KM07}, provides a very clear and transparent way
to exhibit the \textbf{complete resonance set}. The construction is
performed in two steps. First, each \textbf{2D}-vector is regarded
as a node of integer lattice in the spectral space, and nodes
which construct one solution (triad or quartet) are connected. The
result is called \emph{geometrical structure}  of the resonance set
and can be rather nebulous \cite{KM07}. Second, all different
topologically equivalent components of the graph, corresponding to
the geometrical structure, are singled out. The list of all these
elements together with a number, showing how many times a specific
component is met in the solution set, is called \emph{topological
structure} of the resonance set. For instance, in a three-wave resonance
system, most part of the elements are either \emph{isolated triads},
shown as triangles, or \emph{butterflies}, shown as two triangles
with one joint vertex. For dynamical reasons which are explained in the  Sec. \ref{ss:int-clust}, it is important to known whether or not a connection within a cluster is realized via a mode with maximal frequency ($\o_3$-mode). This mode is marked by the letter \emph{A} in each triangle, while two other modes are marked by \emph{P}. Examples of topological elements appearing in
the various water wave systems are shown in Fig.1 in Sec.
\ref{ss:int-clust}. This construction can be extended to the case of four wave resonances but the structure of clusters is more involved \cite{KNR08}.

\section{Dynamics}\label{s:Dyn}

\subsection{Primary clusters}
A more compact form of (\ref{3prim}) can be obtained by a suitable
change of variables and reads
 \be \label{dyn3waves}  \dot{B}_1=   Z B_2^*B_3,\quad
\dot{B}_2=   Z B_1^* B_3, \quad \dot{B}_3= -  Z B_1 B_2. %
\ee 
It has three conservation laws (e.g. \cite{BK09_1})%
\be \label{3ConLawsB}
 I_{23}=|B_2 |^2 + |B_3|^2,  \  I_{13}= |B_1 |^2 + |B_3|^2, \  H_{tr} = \operatorname{Im}(B_1 B_{2}
 B_{3}^*).
\ee %
The first two of them, $I_{23}$ and $I_{13},$  can be rewritten via
energy and enstrophy, while the last one  $H_{tr}$ is the
Hamiltonian of the triad. Notice that though (\ref{dyn3waves}) has 3
complex variables, i.e. 6 real variables, and only three
conservation laws, it is easy to show that it is integrable.
Indeed, one should rewrite it in the standard amplitude-phase
representation $B_j= C_j\exp(i \theta_j)$:
 \bea
\dot{C}_1=Z C_2C_3 \cos \varphi,\quad \dot{C}_2=Z C_1C_3 \cos
\varphi,\quad
\dot{C}_3=- Z C_1C_2 \cos \varphi, \label{Triad-A}\\
\dot{\varphi}\  = -Z\,H_{tr}
(C_1^{-2}+C_2^{-2}-C_3^{-2}).\label{Triad-Ph} \eea
where $\varphi=\theta_1+\theta_2-\theta_3$ is called dynamical
phase. Now the four equations (\ref{Triad-A}),(\ref{Triad-Ph}) have four
real variables, and analytical expressions for amplitudes $C_1, \,
C_2, \, C_3$ are found in Jacobian elliptic functions $\mathbf{dn},
\, \mathbf{sn}, \, \mathbf{cn}$ \cite{Wh37}. Using relations between
squares of elliptic functions, we can represent squared amplitudes as%
 \bea
\label{sol3wavesGeneral}
\begin{cases}
C_1^2(T) = -\mu\,\left(\frac{2 K(\mu)}{Z\,\tau}\right)^2
{{\mathbf{sn}^2}\left(2\,K(\mu )\,
             \frac{{(T-t_0)}}{{\tau}},\mu \right)} + \gamma_1\\
C_2^2(T) =-\mu\,\left(\frac{2 K(\mu)}{Z\,\tau}\right)^2
{{\mathbf{sn}^2}\left(2\,K(\mu )\,
             \frac{{(T-t_0)}}{{\tau}},\mu \right)} + \gamma_2\\
C_3^2(T) = \ \mu\,\left(\frac{2 K(\mu)}{Z\,\tau}\right)^2
{{\mathbf{sn}^2}\left(2\,K(\mu )\,
             \frac{{(T-t_0)}}{{\tau}},\mu \right)} + \gamma_3\,,
\end{cases}
\eea
where $K(\mu)$ is elliptic integral of the first order, and $\mu, \,
t_0, \, \tau, \, \gamma_1,  \, \gamma_2,  \, \gamma_3 $ are explicit
rational expressions of the initial values of the amplitudes
$C_1^2(0), C_2^2(0), C_3^2(0)$ (see \cite{BK09_3}). Since the energy of
each mode $E_{j}$ is proportional to its squared amplitude $C_j^2$
and $0 \le \mathbf{sn}^2 \le 1,$ the characteristic
energy variation of any resonant mode $E_{mode} \sim
\mathbf{sn}^2(k\,t,\mu) $ is bounded: $ 0 \le E_{mode} \le E_0, $
with $E_0=E_0(I_{13}, I_{23}, H_{tr})$ being an explicit function of
the initial conditions.

The dynamical phase  satisfies an evolution equation
(\ref{Triad-Ph}) and its solution \textbf{cannot} be obtained by
simply replacing the solution for the amplitudes in the Hamiltonian
$H_{tr}=C_1 C_2 C_3 \sin \varphi$ and solving for $\varphi$. The
reason is that non-zero $\varphi$ generically evolves between $0$
and $\pi$, crossing the value $\varphi = \pi/2$ periodically. This
implies that $\sin^{-1}$ is double-valued and thus it is not
possible to obtain $\varphi$ in a unique way. Importance of the
dynamical phase is due to the fact that it effects substantially the
magnitudes of $C_1, C_2, C_3$ (see Fig.1 from \cite{BK09_2}).
Solution for $\ph,$ first obtained in \cite{BK09_3}, reads
 \be \label{phase}\varphi(t) = \textrm{sign}
 (\varphi_0)\,\mathrm{\mathbf{arccot}}\left(-\frac{\mu}{|H_{tr}|} \left(\frac{2 K(\mu)}{Z\,\tau}\right)^3
\mathrm{\mathbf{sn\,cn\,dn}} \left(2\,K(\mu )\,
             \frac{{(T-t_0)}}{{\tau}},\mu \right)\right)\,,
\ee%
 with notation  $\mathrm{\mathbf{sn\,cn\,dn}}(x,\mu) \equiv
\mathrm{\mathbf{sn}}(x,\mu) \,\mathrm{\mathbf{cn}}(x,\mu)
\,\mathrm{\mathbf{dn}}(x,\mu)\,. $

Substantially less is known about the integrability of (\ref{4prim}).
Its analysis can probably be carried out along the same lines as for
(\ref{3prim}). A promising start has been made in \cite{SS05}, though
analysis was not brought all the way to explicit formulae
similar to those obtained for (\ref{3prim}). Also the importance of
the dynamical phase has not been put to work yet, while (\ref{4prim}) is
analyzed in the complex form and the effect of the dynamical phase is
hidden. For the general form of a quartet and for arbitrary initial
conditions, (\ref{4prim}) "does not exhibit strict periodicity"
\cite{SS05}, though in the particular case of a trident, its
behavior is periodic \cite{SS85} (both statements refer to the
results of numerical simulations).

\begin{table}[htp]
\begin{tabular}{|c|c|c|}
  \hline 
   1 & 2 & 3 \\ \hline
\hline
     &  &   \\
   $\sigma k^{3/2} $ & $V^3_{12}$ &  $\quad{i \over 8 \pi \sqrt{2 \sigma}} \sqrt{\omega_{1}
\omega_2 \omega_3} \times  \left[ {\mathcal{K}_{k_2, k_3} \over
k_1\sqrt{ k_2 k_3 } } -
 {\mathcal{K}_{k_1, -k_2} \over  k_3 \sqrt{ k_1 k_2 } }
-
 {\mathcal{K}_{k_1, -k_3} \over k_2 \sqrt{ k_1 k_3 }  }
\right]\, $
  \\
      & \cite{ZF67} &   \\
     &  &
 where
 $
 \mathcal{K}_{k_2, k_3}  = ({ \mathbf{k}}_2  \cdot {\mathbf{k}}_3) +  k_2 k_3\ $.
   \\
     &  &   \\
\hline
     &  &    \\
  $ (g \, k + \sigma k^3)^{1/2}$ & $V^3_{12}$ &   $\quad \quad  \quad \frac{(\o_2^2-\o_2 \o_3 + \o_3^2)}{\o_1}k_1 -\o_2 k_2  + \o_3 k_3 $  \\
           & \cite{McG65} &   \\
    &  & $+ ({\mathbf{k}}_2  \cdot {\mathbf{k}}_3) \big( \frac{\o_2 }{k_2}- \frac{\o_3}{k_3} - \frac{\o_2 \o_3}{\o_1} \frac{k_1}{k_2 k_3}\big)$  \\
       &  &   \\
\hline
     &  &     \\
  $g k^{1/2} $ & $ T^{12}_{34}$ & $\quad \quad  \frac{1}{16\pi^2 (k_1k_2k_3k_4)^{1/4}}\Big[-12k_1k_2k_3k_4$  \\
    & \cite{Z99} &    \\
     &  & $-2 \frac{(\o_1+\o_2)^2}{g^2} [\o_3 \o_4({\mathbf{k}}_1  \cdot {\mathbf{k}}_2 -  k_1 k_2)+ \o_1 \o_2({\mathbf{k}}_3  \cdot {\mathbf{k}}_4 -  k_3 k_4)]$  \\
         &  &   \\
     &  & $-2 \frac{(\o_1-\o_3)^2}{g^2} [\o_2 \o_4({\mathbf{k}}_1  \cdot {\mathbf{k}}_3 +  k_1 k_3)+ \o_1 \o_3({\mathbf{k}}_2  \cdot {\mathbf{k}}_4 +  k_2 k_4)]$   \\
         &  &    \\
     &  & $-2 \frac{(\o_1-\o_4)^2}{g^2} [\o_2 \o_3({\mathbf{k}}_1  \cdot {\mathbf{k}}_4 +  k_1 k_4)+ \o_1 \o_4({\mathbf{k}}_2  \cdot {\mathbf{k}}_3 +  k_2 k_3)]$   \\
         &  &    \\

     &  & $+4 (\o_1+\o_2)^2\frac{({\mathbf{k}}_1  \cdot {\mathbf{k}}_2 -  k_1 k_2)(-{\mathbf{k}}_2  \cdot {\mathbf{k}}_4 +  k_2 k_4)}{\o^2_{1+2}-(\o_1+\o_2)^2} $   \\
         &  &    \\
     &  & $+4 (\o_1-\o_3)^2\frac{({\mathbf{k}}_1  \cdot {\mathbf{k}}_3 +  k_1 k_3)({\mathbf{k}}_2  \cdot {\mathbf{k}}_4 +  k_2 k_4)}{\o^2_{1-3}-(\o_1-\o_3)^2}$   \\
         &  &    \\
     &  & $+4 (\o_1-\o_4)^2\frac{({\mathbf{k}}_1  \cdot {\mathbf{k}}_4 +  k_1 k_4)({\mathbf{k}}_2  \cdot {\mathbf{k}}_3 +  k_2 k_3)}{\o^2_{1-4}-(\o_1-\o_4)^2}$  \\
         &  &    \\
     &  &  $+({\mathbf{k}}_1  \cdot {\mathbf{k}}_2 +  k_1 k_2)({\mathbf{k}}_3  \cdot {\mathbf{k}}_4 +  k_3 k_4)$  \\
     &  &    \\
     &  &  $ +(-{\mathbf{k}}_1  \cdot {\mathbf{k}}_3 +  k_1 k_3)(-{\mathbf{k}}_2  \cdot {\mathbf{k}}_4 +  k_2 k_4) $  \\
     &  &    \\
     &  & $+(-{\mathbf{k}}_1  \cdot {\mathbf{k}}_4 +  k_1 k_4)(-{\mathbf{k}}_2  \cdot {\mathbf{k}}_3 +  k_2 k_3)\Big]$   \\
     &  &    \\
  \hline 
\end{tabular}
\vspace{1cm} \caption{Overall data on three types of water waves:
 form of dispersion function,
notation for coupling coefficient and original reference, form of
coupling coefficient (1st, 2d and 3d columns correspondingly).}
\end{table}%

\subsection{Clusters of two and more triads}\label{ss:int-clust}

\begin{figure}[h]
\includegraphics[width=9.5cm,height=5cm]{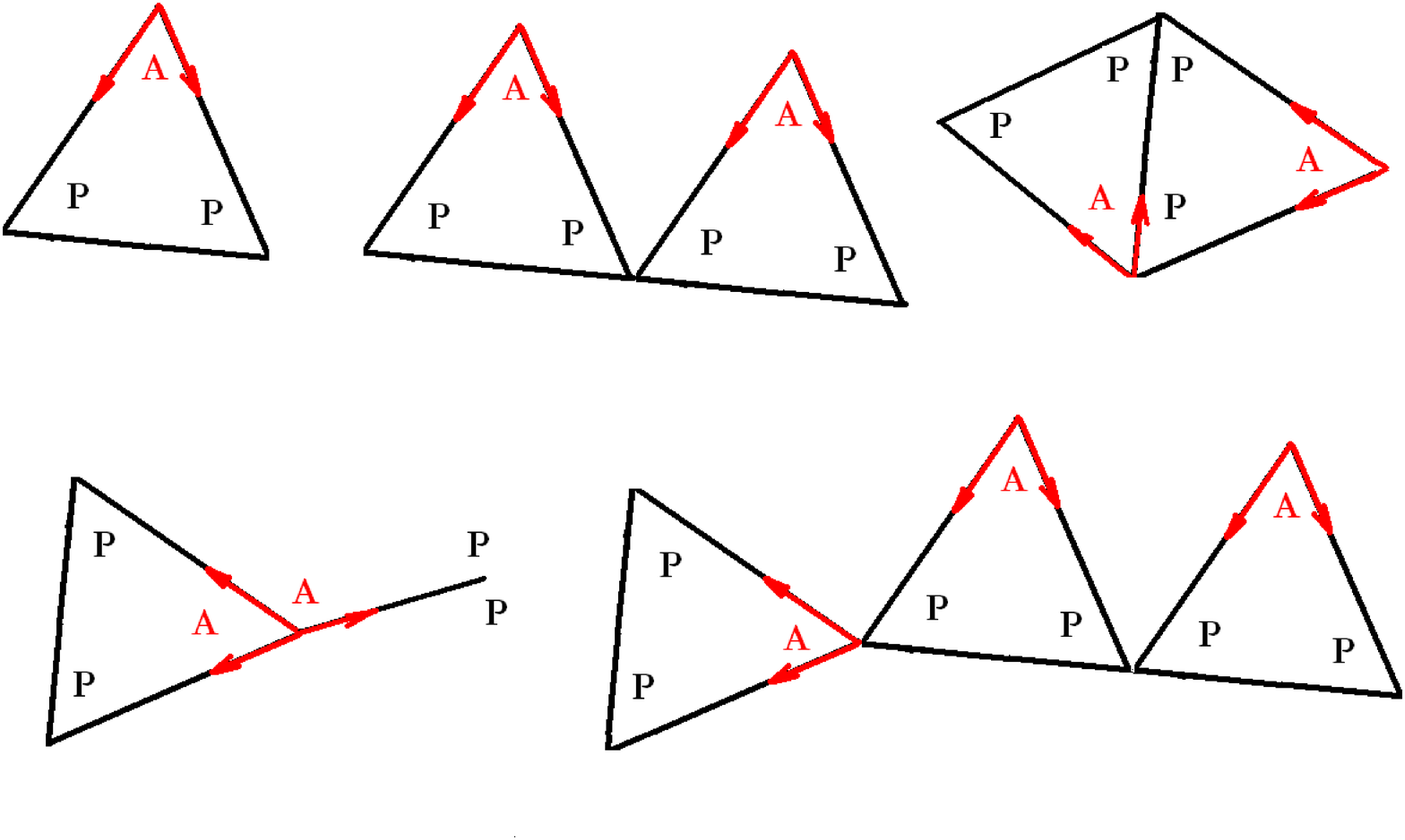}
\vspace{-1.5cm} \caption{\label{f:top-struc}  Triad with marked
vertices; PP-butterfly; AP-PP-kite, AA-ray and AP-PP-chain of three
triads.}
\end{figure}

Dynamics of a cluster consisting of a few connected triads depends
on whether or not the connecting mode is the mode with maximal
frequency $\o_3$ in one or both triads \cite{KL08}. It follows from
the form of conserved quantities $I_{23}, I_{13}$ (see
 (\ref{3ConLawsB})) that if only one of the  modes with
 lower frequencies $\o_1$ and $\o_2$ is exited, it can not share its
 energy with other modes. Therefore these two modes are called
 \emph{P-modes}, \emph{P }from passive. On the contrary, $\o_3$-mode is called \emph{A-mode}, \emph{A} from active, because
 being exited initially, it causes exponential growth of P-modes
 amplitudes, until all the modes will
have comparable magnitudes of amplitudes (cf. the criterium of
instability in \cite{Has67}). In Fig. \ref{f:top-struc} examples of
simple resonance clusters are given, with marked vertexes. Each
isolated marked graph defines uniquely some dynamical system. For
instance, a PP-butterfly consisting of triad $a$ and triad $b$,
connected via the mode $B_{1a}=B_{1b},$ is covered by
 \bea \label{PP}
\dot{B}_{1a}=  Z_a B_{2a}^*B_{3a} +   Z_b B_{2b}^*B_{3b}, \quad
\dot{B}_{2a}=  Z_a B_{1a}^* B_{3a}, \nonumber\\ \dot{B}_{2b}= Z_b
B_{1a}^* B_{3b}, \quad \dot{B}_{3a}=  - Z_{a} B_{1a} B_{2a}, \quad
\dot{B}_{3b}=  - Z_{b} B_{1a} B_{2b}\ .
 \eea%
and conservations laws have the form
 \bea \label{int-PP}
I_{23a}=|B_{2a} |^2 + |B_{3a}|^2 \,, \quad
 I_{23b}= |B_{2b} |^2 + |B_{3b}|^2 \,, \nonumber \\
 I_{ab}= |B_{1}|^2 + |B_{3a} |^2 + |B_{3b}|^2 \,, \quad
H_{but} = \operatorname{Im}(Z_a B_1 B_{2a} B_{3a}^* + Z_b B_1 B_{2b}
B_{3b}^*).\nonumber 
  \eea
 \vspace{-0.5cm}

Some scattered results on the integrability of resonance clusters can
be found in \cite{BK09_1,BK09_3,Ver68a,Ver68b}. For instance,
PP-butterfly (\ref{PP}) is known to be integrable if 1) $Z_a/Z_b = 1$ or 2 or 1/2; and 2)
$Z_a, Z_b$ are arbitrary and $H_{but}=0$. However, it is not proven that this list is exhaustive. Some of the kites, rays
and finite chains of triads (see Fig.\ref{f:top-struc}) are
integrable, but complete classification of integrable resonance
clusters is still to be constructed.

No quantitative results are known about dynamics of resonance clusters in
four-wave systems. Simple qualitative scenario suggested in \cite{KNR08} is  based on the pure kinematic considerations  -- reservoirs formed by many angle quartets in quasi-thermal equilibrium with sparse links between them formed by scale-resonances -- and has yet to be justified. At least, magnitudes of coupling coefficients have to be computed; this qualitative scenario will be invalidated, for instance, if most of angle-resonances have zero coupling coefficients, or their magnitudes are substantially smaller then those of scale-resonances, etc.

\begin{remark}
 Three-wave system might possess a four-wave resonance cluster but it will not be a primary cluster (see Fig.\ref{f:top-struc}, kite or ray).
This can easily be seen from the form of the corresponding dynamical system.
For instance, a PP-PP kite, with
 $ B_{1a}=B_{1b} $ and $ B_{2a}=B_{2b},$ has dynamical system 
\bea \label{PP-PP}
\dot{B}_{1a}=  B_{2a}^* (Z_a B_{3a} +   Z_b B_{3b}) \,, \quad
\dot{B}_{2a}=  B_{1a}^* (Z_a B_{3a} +  Z_b B_{3b}) \,, \nonumber \\
\dot{B}_{3a}=  - Z_{a} B_{1a} B_{2a} \,,\    \dot{B}_{3b}=  - Z_{b}
B_{1a} B_{2a}\ ,
 \eea%
which is different from (\ref{4prim}). Clusters of this form are indeed observed among oceanic planetary waves \cite{KRFMS08} and their integrability is proven in \cite{BK09_3}.
\end{remark}

\begin{remark} All the results above have been obtained for water waves without
vorticity. As has been shown in \cite{CK09}, non-zero vorticity can
generate non-linear resonances. It would be interesting to
 derive dynamical
equations for resonance clusters in this case. The presence of
non-zero vorticity invalidates the existence of a velocity potential
for the flow and harmonic function theory is not readily available
for the analysis. On the other hand, both multi-scale methods and
variational techniques can also be used for the case of non-zero
vorticity \cite{CSS,CS2,Jo,W2}.
\end{remark}

\section{Numerical simulations and laboratory experiments}

In the past, numerical simulations \cite{AS06,LNP06,PZ00,ZKPD05} and
laboratory experiments  \cite{DLN07,FLF07} have been mostly
performed for checking the prediction possibilities of classical wave
turbulence  theory with continuous wave spectra (CWT), that is, kinetic equations and power energy
spectra. It has been established that the so-called infinite-box limit
assumed by CWT theory is not achieved, and even in a 10m x 6m laboratory
flume finite-size effects are very strong \cite{DLN07}.

Very interesting results are presented in \cite{CHS96}, where chains
of three connected quasi-resonant triads of gravity-capillary waves
have been identified in the experimental data, dynamical system
consisting of 7 nonlinear ODEs was written out explicitly and solved
numerically. Experimental data have been compared with the numerical
predictions of 7-modes model computed for constant dynamical
phase, $\ph=\pi.$  Predictions were in qualitative but not in
quantitative agreement with the experiment
(magnitudes of amplitudes were underestimated
by the theoretical prediction). Quite recently results of
laboratory experiments with surface waves on deep water were
reported \cite{HPS06,SH07} in which regular, nearly permanent
patterns on the water surface have been observed. It would be
interesting to attribute these regular patterns to specific
resonance clusters, like it has been done in \cite{CHS96} but taking into account the dynamical phase also.

In this context, we would like to mention results presented in \cite{RS09}. Pattern formation due to nonlinear resonances is studied numerically for a somewhat simplified model PDE (nonlinear terms do not include any derivatives). Dissipation and multifrequency forcing are taken into account; the importance of the modes' phases is fully recognized; pattern-forming modes are computed via fast Fourier transform. The main drawback of this Ansatz is that the model equation, though it has a Hamiltonian limit, can not be derived from Navier-Stokes equation and its dispersion function is different from the real-world case. The choice of the model PDE has the purpose to make numerical simulations less time-consuming.  On the other hand, methods developed in \cite{KK06-1,KK06-2,KK06-3} allow naturally direct computation of pattern-forming modes without performing any numerical simulation with nonlinear PDEs.

The knowledge of the structure of resonance clusters might shed some
new light into the origin of such  well-known physical phenomena as
Benjamin-Feir instability, \cite{BF67}.  As it was shown recently
in~\cite{HH03,HHS03,StabilBF}, its explanation as the modulational
instability, though well established in water waves theory, has to
be seriously reconsidered for it over-predicts the growth rate of the
waves. Another inconsistency is due to the fact that it can easily be stabilized by arbitrary small dissipation. The other way to treat the problem would be to  explain the modulational instability in terms of non-collinear
resonances.

\subsection{Quasi-resonances and approximate interactions}\label{s:Quasi}

\begin{figure}[h]
\includegraphics[width=6.0cm,height=4.0cm]{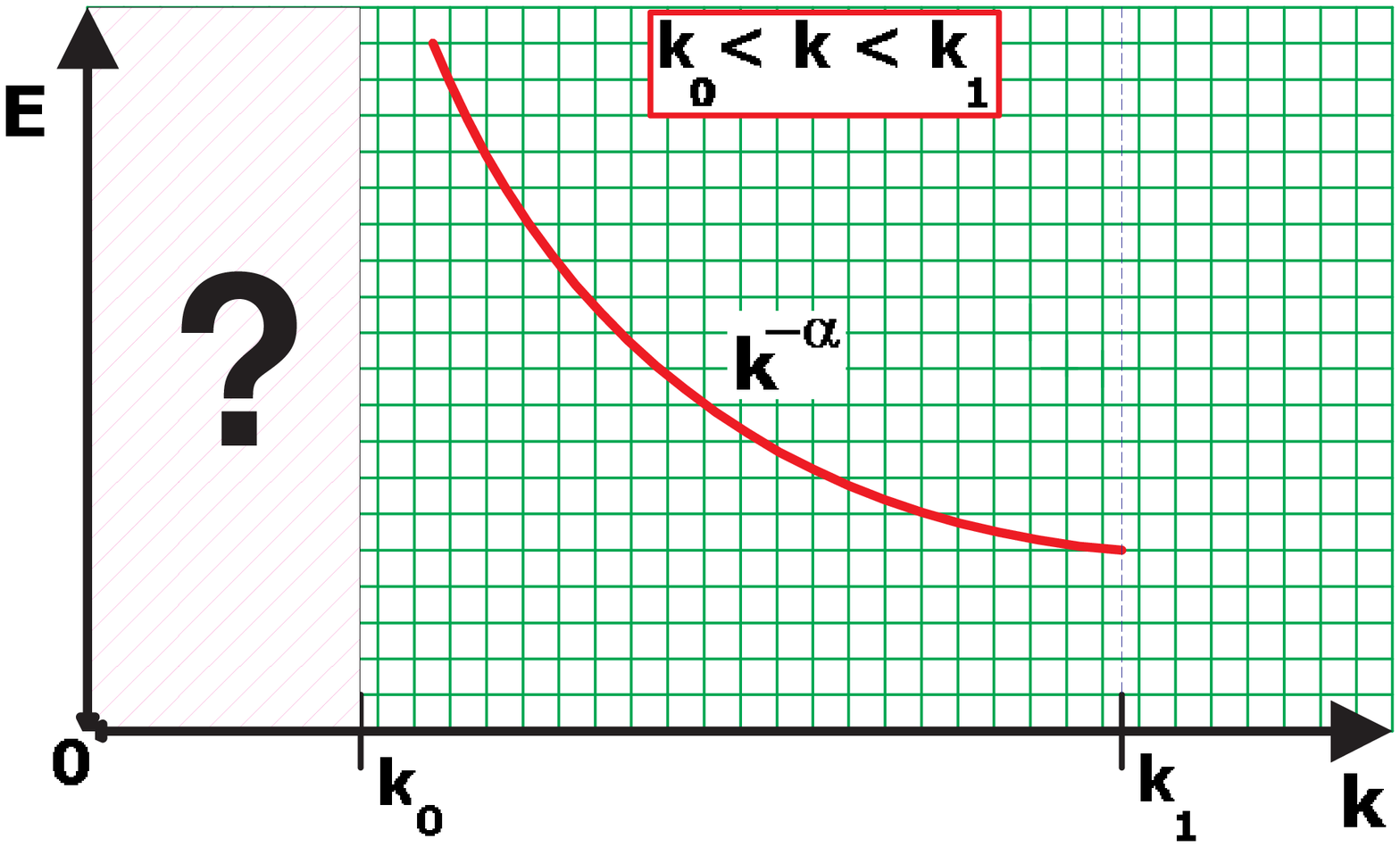}
\includegraphics[width=6.0cm,height=4.0cm]{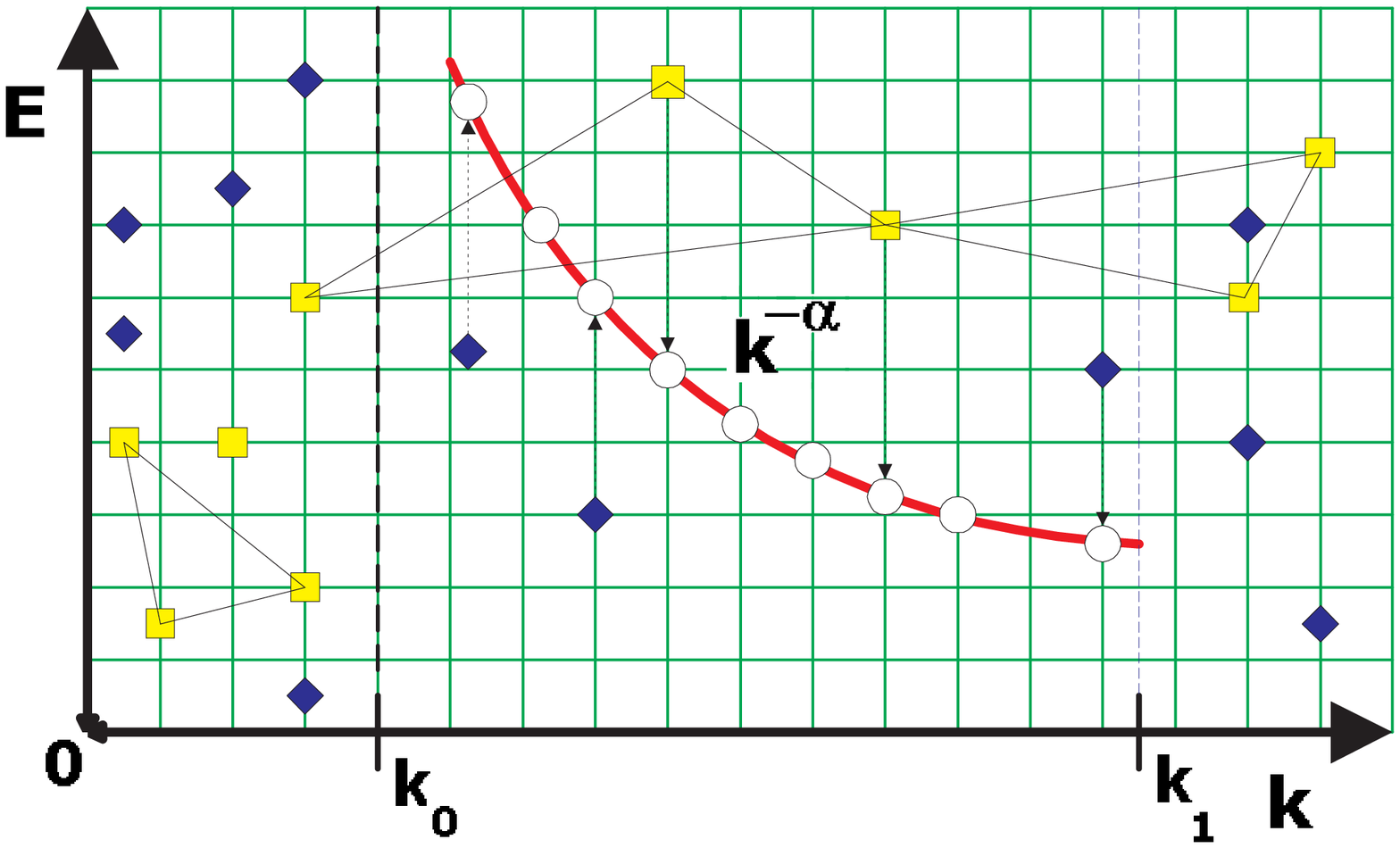}
\caption{\label{f:LamTur}}
\end{figure}

The
reason why the predictions of CWT theory  are mostly not corroborated, both in numerical simulations and laboratory experiments, is that resonance broadening $\Omega$ (also called resonance width or frequency discrepancy or frequency mismatch etc.), defined as
\be \label{Width} |\o_1 \pm \o_2 \pm \cdots \pm \o_s| = \Omega >0,\quad
\mathbf{k}_1 \pm \mathbf{k}_2 \pm \cdots \pm  \mathbf{k}_s=0, \, s < \infty \,,    \ee%
is \emph{not big enough} for CWT theory to be
applicable. CWT theory is supposed to work when the
resonance broadening $\Omega$ is greater than the spacing
$\delta_\omega$ between the adjacent wave modes \be
 \Omega > (\p \o
/\p k) 2 \pi/ L,
\label{quasi-r}
\ee where $L$ is the box size.
The trouble comes from the notorious small divisor problem known from KAM-theory;  in CWT theory it appears first in \cite{ZSh90}, Eq.(2.5.2). The CWT theory
assumes weak nonlinearity, randomness of phases, infinite-box limit, and existence of the inertial interval
$(k_o, k_1)$, where energy input and dissipation are balanced. Under these assumptions, the wave system is energy conserving, and
wave kinetic equations describing the wave spectrum and energy power spectra $k^{-\a}, \, \a>0, $ have stationary solutions (\cite{Ph60,ZF67,ZLF92}, etc.) For $k\ll k_0$, so-called finite-size effects take place which are due to boundary conditions and should be regarded separately. For $k\gg k_1,$ dissipation suppresses nonlinear dynamics (see  Fig.\ref{f:LamTur}, left).
In the last decade, this standard view proved to be incomplete while finite-size effects are well observable within inertial interval; this led first to introducing such special types of wave turbulence as frozen turbulence of capillary waves \cite{PZ00} and mesoscopic turbulence of gravity waves \cite{ZKPD05}. Quite recently, a novel two-layer model of laminated turbulence has been developed \cite{K06-1}. It
allows to explain finite-size effects in the same frame: each small divisor leaves a gap in the power spectrum, shown as an empty circle in Fig.\ref{f:LamTur} on the right. The lower boundary for the radius $\mathcal{R}$ of each circle can be computed using Thue-Siegel-Roth theorem \cite{K07} for the case of an irrational dispersion function, which covers all water waves (\ref{CapDeepWat})-(\ref{GrDeepWat}). The gaps corresponding to the exact and quasi-resonances are shown by yellow squares; their dynamics is covered by the dynamical systems of corresponding resonance clusters (\cite{K94}, Fig.1). Some gaps, shown by blue diamonds, correspond to the discrete modes that do not take part in resonances; they just keep their initial energies (\cite{K94}, Fig.3).
\bigskip
\begin{definition}
Solutions of (\ref{Width}) are called:
\emph{exact resonances}, if $ \Omega =0;$
\emph{quasi-resonances}, if $$ 0< \mathcal{R} \le \Omega \wt{<} (\p \o
/\p k) 2 \pi/ L;$$ and
\emph{approximate interactions}, if $$\Omega \gg (\p \o
/\p k) 2 \pi/ L.$$
Exact resonances and quasi-resonances form \emph{discrete layer} of turbulence, while approximate interactions form \emph{continuous layer}. Correspondingly, the notion of discrete wave turbulence (DWT) is nowadays used, as opposed to CWT.
\end{definition}
\bigskip

\begin{remark}
Previous to our theory \cite{K06-1}, the standard counterpart of CWT were
low-dimensioned systems characterized by a \emph{small number of mode}s included. On the other hand, DWT is characterized by the \emph{clustering} itself, and \emph{not by the number of modes} in particular clusters which can be fairly big.
\end{remark}

The  minimal resonance broadening, necessary for quasi-resonances to appear, has been investigated numerically in \cite{TY04} both for capillary and gravity water waves. Numerical simulations \cite{Tan01,Tan07,TY04} showed that exact and quasi-resonances among the gravity waves are observable not on the time scale $\mathcal{O}(t/\e^4)$ of kinetic theory but on the linear time scale. This means that dynamics of discrete layer but not the kinetic equation should be used as the base for a short-term forecast.
The classical notion of wave interactions would be quite inappropriate -- even misleading -- for the description of all these important effects.

\section{Summary}

In 1981, in the paper with the title "Wave interactions -- evolution
of the idea," \cite{Ph81}, O. Phillips wrote prophetical words: "New
physics, new mathematics and new intuition is required" in order to
gain some understanding of finite-size effects in wave turbulent systems.
 Few years later, two  -- nowadays classical --
books were published: "Wave interactions and fluid flows,"
\cite{Cr85},  and "Kolmogorov spectra of turbulence,"\cite{ZLF92}.
Appearing in 1985, the first one gives comprehensive account of the
theory and experiments on exact resonances,
 quasi-resonances and approximate interactions -- all together named \emph{wave interactions} - of primary clusters
 in three- and four-wave systems. The second book came to world in 1992,
 with a beautiful statistical theory of wave turbulence  which is
 due to the \emph{exclusion of exact resonances and quasi-resonances} from
 the consideration. The first attempt to fill in the gap between
these two treatises has been made in 1994, in a small paper
"Weakly nonlinear theory of finite-size effects in
 resonators", \cite{K94}. Today, some 15 years later, we
 know much more about these effects and can formulate the
three most important mathematical problems whose solutions would
contribute enormously to our understanding of nonlinear
resonance dynamics of water waves.

\emph{1st problem.} Given a dispersion function and a small parameter,
by some standard technique initial nonlinear PDE with fixed
boundary conditions can be transformed into a set of dynamical
systems corresponding to resonance clusters. Each of them is a
Poincare normal form. Classification of all normal forms appearing this way
according to their integrability properties is needed.

\emph{2nd problem.} Wave turbulence with continuous spectra is
described quite comfortably by the energy power spectra $k^{-\a}$,
stationary solutions of kinetic equations: the only variable here is the
wave length $k,$ while $\a$ is a parameter, defined by the wave type. Only the fact that the wave system has a conservation
law, say its energy $E$, is used, not the value of $E.$
The finite answer is therefore independent on the initial conditions. The situation is
different for discrete wave turbulence. Energy transfer is presently described by
exact formulae (\ref{sol3wavesGeneral}) which are definitely not
easy to use while initial values of the conservation laws
(\ref{3ConLawsB}) are included explicitly and they define
 the period of energy oscillation of each resonant mode. Some simplified description is needed, perhaps also
stationary, with $\ph$ regarded as a (different) constant at each
moment of time.

\emph{3rd problem.} To describe the transition from discrete to
continuous layer of turbulence in terms of energy flow is a very challenging
problem, which probably can not be solved before a simplified
description of the discrete layer will be obtained.

\medskip
Received xxxx 20xx; revised xxxx 20xx.
\medskip

\end{document}